\documentclass[titenlines, reprint, twocolumn, pra,floatfix,groupedaddress,superscriptaddress,aps]{revtex4}
\usepackage{graphicx}
\usepackage{rotating}
\usepackage{array}
\usepackage{multirow}
\usepackage{epstopdf}
\usepackage{amsmath}
\usepackage{amsfonts}
\usepackage{setspace}
\usepackage{braket}
\usepackage{color}

\usepackage{float}

\usepackage{hyperref}

\DeclareMathOperator{\Tr}{Tr}

\begin{document}
\title{Homotopy continuation method for solving Dyson equation fully self-consistently: theory and application to NdNiO$_2$} 

\author{Pavel Pokhilko}
\email{pokhilko@umich.edu}
\affiliation{Department  of  Chemistry,  University  of  Michigan,  Ann  Arbor,  Michigan  48109,  USA}
\affiliation{Current address: Department of Physics and Astronomy, Purdue University, West Lafayette, Indiana 47907, USA}
\author{Dominika Zgid}
\affiliation{Department  of  Chemistry,  University  of  Michigan,  Ann  Arbor,  Michigan  48109,  USA}
\affiliation{Department of Physics, University of Michigan, Ann Arbor, Michigan 48109, USA }

\renewcommand{\baselinestretch}{1.0}
\begin{abstract}
Solution of the Dyson equation for the small-gap systems can be plagued by large non-converging iterations. 
In addition to the convergence issues, due to a high non-linearity, the Dyson equation may have multiple solutions. 
We apply the homotopy continuation approach to control the behavior of iterations. 
We used the homotopy continuation to locate multiple fully self-consistent GW solutions for NdNiO$_2$ solid 
and to establish the corresponding Hartree--Fock limits. 
Some of the solutions found are qualitatively new and help to understand the nature of
electron correlation in this material. 
We show that there are multiple low-energy charge-transfer solutions leading to a formation of charge-density waves. 
Our results qualitatively agree with the experimental conductivity measurements. 
To rationalize the structure of solutions, we compare the k-point occupations
and generalize the concept of natural difference orbitals for correlated periodic solids.  
\end{abstract}
\maketitle

\section{Introduction}
One-particle Green's function theories provide a powerful and rigorous theoretical framework \cite{Mahan00,Negele:Orland:book:2018,Martin:Interacting_electrons:2016} for
describing ionization and electron attachment processes 
and giving access to one-particle and two-particle observables, magnetic properties, geometries, 
and thermodynamic quantities at a finite temperature.  
Self-consistent Green's function theories that solve the Dyson equation fully self-consistently until convergence is reached guarantee conservation of the number of particles, momenta, 
and angular momenta \cite{Baym61,Baym62}.  
Additionally, at full self-consistency ``memory'' of the starting point is removed, making such calculations 
independent of starting orbitals.
A subset of such Green's function theories, called $\Phi$-derivable theories \cite{Baym61,Baym62}, 
additionally guarantees conservation of energy, 
thermodynamic consistency, current continuity, and gauge invariance of the Luttinger--Ward functional.  
Many of these properties of Green's function methods are crucial for practical applications to enable predictions that closely match results from advanced wave-function-based theories and experimental observations\cite{Wen:2023,Abraham:X2C-GW:2024,Yeh:X2C:GW:2022,Pokhilko:local_correlators:2021,Pokhilko:Neel_T:2022,Harsha:GW:challenges:2024}.  

Despite a very significant progress in improving numerical aspects of  Green's function theories such as increased efficiency due to better frequency grids\cite{Kananenka:grids:2016,Kananenka16,Iskakov_Chebychev_2018,dong2020legendrespectral,Yoshimi:IR:2017}, 
lowered scaling due to integral compression and decompositions\cite{Rusakov14,Rusakov:SEET:2019,Tran_generalized_seet,Iskakov20,Yeh:GPU:GW:2022,Yeh:X2C:GW:2022,Yeh:THC-GW:2024,Yeh:THC-RPA:2023,Pokhilko:THC-GWSOX:2024,Pokhilko:THC-G3W2:2025},  
optimized computer implementations\cite{Yeh:GPU:GW:2022,Green:2025,Yeh:THC-GW:2024,Yeh:THC-RPA:2023}, 
and advanced convergence algorithms\cite{Pokhilko:algs:2022}, 
a number of mathematical and theoretical aspects of Green's function methods still remain unexplored.

Due to a high nonlinearity, the Dyson equation allows us to obtain multiple solutions\cite{Mochena:Fukutome:broken_symmetry:GF,Pokhilko:local_correlators:2021,Pokhilko:THC-G3W2:2025}. 
For insulators, many of such solutions have a well-understood character and physical meaning (e.g., triplet and broken-symmetry solutions) and the iterative procedure can be converged to them by providing a starting point that can be easily rationalized and understood.
Previously, we found such solutions for molecules\cite{Pokhilko:local_correlators:2021,Pokhilko:THC-G3W2:2025} 
and insulating solids\cite{Pokhilko:BS-GW:solids:2022,Pokhilko:NO:correlators:2023,Pokhilko:Neel_T:2022} and successfully used them to 
evaluate effective exchange Heisenberg couplings through the broken-symmetry approach and 
even Neel temperatures\cite{Pokhilko:Neel_T:2022}.  

However, for the small-gap materials, there are several complications: (i) it is not apparent which solutions exist since the number of possible solutions that are very close in energy can be huge; (ii) the capture of a particular solution can be numerically difficult due to poorly converging or even non-converging iterations; (iii) the self-consistency cycle can oscillate between solutions; (iv) even if the convergence is reached, the solution found  may not be the one that is desired, possibly causing mistakes in the interpretation of such calculations.  

While in the zero-temperature wave-function Hartree--Fock formulation there are approaches partly 
solving some of these issues 
(e.g., the maximum overlap method\cite{Gill:MOM:08} during iterations that can provide occupations closest to the targeted ones),   
these approaches cannot be easily generalized to the finite-temperature correlated case.  
In this paper, we introduce a new convergence technique using the homotopy continuation.  
Using this technique, we show how the aforementioned issues can be solved at least partly. 

In particular, the homotopy continuation establishes 
the correspondence between the mean-field solutions
and the solutions of the 
correlated Dyson equation. We apply this technique within the fully self-consistent GW solutions applied to 
to NdNiO$_2$ solid---a correlated quantum material that has been a subject of several controversies over the last years, such as whether NdNiO$_2$ has the same type of electronic structure as infinite-layer cuprates, whether NdNiO$_2$ has a charge-transfer order, or whether it is a Mott insulator. These controversial findings are present within both the experiment and the theory. For example, two RIXS experiments were interpreted as the evidence of charge order\cite{Krieger:NdNiO2:RIXS:2022,Zhou:CDW:NdNiO2:2022}, while the subsequent RIXS experiment\cite{Shen:noCDW:NdNiO2:2024} was interpreted as an absence of the charge order. Among the theoretical publications, multiple models with adjustable parameters \cite{Chang:NdNiO2:2020,Wu:twoband:NdNiO2:2019,Vishwanath:tJ:NdNiO2:2020,Sawatzky:NdNiO2:model:2020,Stepanov:NdNiO2:2024,Oles:NdNiO2:s-orbs:2022,Oles:NdNiO2:VCA:2025} have been constructed predicting different physics. The derivation of these models often relies either on another model with adjustable parameters, such as the Hubbard model, or on DFT, which is known to give a qualitatively wrong description of strong correlation.

Using the homotopy, we try to explain the numerical origin of some of these controversies. To do so, we successfully locate multiple solutions and explore their uniform and charge-transfer characters. 
The uniform solutions have the same number of electrons of a given spin for every k-point.
The charge-transfer solutions have different number of electrons for different k-points. One can imagine that these charge-transfer solutions arise as a process of charge (electron) transfer between different k-points of the uniform solution.
Our results shed some light on previously reported controversial findings for these different solutions.

In order to distinguish different solutions, we establish a series of unique identifiers by generalizing the concept of natural difference orbitals (NDOs) to solids. These identifiers are then used for analyzing the physical meaning of the solutions found.
Our numerical results also indicate the presence of instabilities of several types.

The imaginary-time one-particle Green's function is defined as
\begin{gather}
G_{\textbf{pq}} (\tau) = -\frac{1}{Z} \Tr \left[e^{-(\beta-\tau)(\hat{H}-\mu \hat{N})} \mathbf{\hat{p}} e^{-\tau(\hat{H}-\mu \hat{N})} \mathbf{\hat{q}}^\dagger  \right], \\
Z = \Tr \left[ e^{-\beta(\hat{H}-\mu \hat{N})} \right], 
\end{gather}
where $Z$ is the grand canonical partition function, 
$\mu$ is the chemical potential, 
$\tau$ is the imaginary time, 
$\beta$ is the inverse temperature, 
$\hat{N}$ is the particle-number operator, 
$\mathbf{\hat{p}}$ and $\mathbf{\hat{q}}^\dagger$ are annihilation and creation operators, respectively, 
expressed in a momentum-dependent spin-orbital basis. 

We define the Green's function of independent electrons $G_0$ as
\begin{gather}
{G}^{-1}_0(i\omega_n) = i\omega_n +\mu \hat{N} - \hat{H}_0, \\
\omega_n = \frac{(2n+1) \pi}{\beta}, 
\end{gather}
where $\hat{H}_0$ is the Hamiltonian of independent electrons and
$\omega_n$ is the fermionic (odd) Matsubara frequency. 

The Dyson equation is written as
\begin{gather}
G^{-1}(i\omega_n) = G_0^{-1}(i\omega_n) - \Sigma[G](i\omega_n),  \\
\Sigma[G](i\omega_n) = \Sigma_\infty[G] + \Sigma_{dyn}[G](i\omega_n),
\end{gather}
where $\Sigma[G]$ is the self-energy which is a functional depending only on the full Green's function $G$ 
and two-body integrals. 
The static (frequency-independent) part of the self-energy is $\Sigma_\infty$ and 
$\Sigma_{dyn}$ is the dynamical part of the self-energy. 
Approximate fully self-consistent Green's function methods are approximating of the functional dependence $\Sigma[G]$. 
The simplest approximation, called the grand-canonical Hartree--Fock approximation, is constructed only from the static part of the self-energy, 
$\Sigma[G] \approx \Sigma_\infty[G]$.  
In the zero-temperature limit, such an approximation often coincides with the more commonly known wave-function Hartree--Fock approximation. 
However, the Green's function formulation of the Hartree--Fock approximation may not preserve identpotency of the one-particle density matrix and can lead to solutions that cannot be represented as a single determinant\cite{Pokhilko:tpdm:2021}.  

Correlated theories approximate the dynamical part of the self-energy. 
In this work, we consider the GW approximation\cite{Hedin65,G0W0_Pickett84,G0W0_Hybertsen86,GW_Aryasetiawan98,Stan06,Koval14,scGW_Andrey09,GW100,Holm98,QPGW_Schilfgaarde,Kutepov17,Iskakov20,Yeh:X2C:GW:2022,Yeh:GPU:GW:2022}, 
which is the first perturbative order in the \emph{screened} interaction $W$.  
The GW approximation is the simplest fully self-consistent approximation of 
the Hedin's coupled system of equations\cite{Hedin65}. 
The details of the implementation used here as well as algebraic expressions are available in Ref.\onlinecite{Iskakov20,Yeh:GPU:GW:2022}.  

\section{Homotopy}\label{sec:homotopy}
Homotopy is defined as a continuous map between topological spaces
\begin{gather}
h: X \times [0,1] \rightarrow X,
\end{gather}
such that $h(x,0) = f(x)$ and $h(x,1) = r(x)$ 
are continuous functions. Homotopy defines a continuous deformation of $f(x)$ into $r(x)$. 
The simplest homotopy is linear homotopy $h_\lambda(x) = (1-\lambda) f(x) + \lambda r(x)$, 
which we actively use in this work. 

In quantum chemistry, linear homotopy was used for geometry interpolation\cite{Kaiser:Chiral:18}, 
finding multiple Hartree--Fock solutions\cite{Kowalski:multiple:CCD:1998}, 
multiple coupled-cluster solutions\cite{Kowalski:multiple:CCD:1998,Jankowski:multiple:CC:1994,Faulstich:geom:CC:2024,Kowalski:MM:CC,Faulstich:homotopy:CC:2023,Faulstich:CC:2024,Csirik:CC:revisited:2023}, 
scaling of the effective action\cite{Kozik:homotopy:2021}, 
and estimates of the number of solutions of the Hedin's system of equations\cite{Gross:GW:multiplicity:2015}.

Homotopy is useful not only for topology\cite{Arkowitz:homotopy:2011}, the numerical solution of differential equations\cite{Liao:1999,Allgower:2012}, 
search of fixed points\cite{Chow:1978,Allgower:1993}, and for
numerical search of roots of polynomial systems of equations\cite{Morgan:all_poly:1987a,Morgan:all_poly:1987b,Durand:homotopy:1998}. Linear homotopy continuation is especially powerful for solving polynomial systems 
because of its capacity to find all complex roots. 
In practice, however, there could be issues with numerical precision as well as with cases 
where the rank of the Jacobian is not full, preventing a smooth continuation. 
To mitigate such issues, 
$R(x)=0$ a polynomial system is replaced with $e^{i\phi} R(x)=0$ a system that has the same roots. 
This results in a family of homotopies depending on a parameter $\phi$.
\begin{gather}
H_\lambda^{\phi}(x) = (1-\lambda)F(x) + \lambda e^{i\phi} R(x).
\end{gather}

For strongly correlated or even moderately correlated systems, the magnitude of the self-energy is large. 
If the initial approximation is based on the Hartree--Fock or density functional theory, 
the Green's function iteration often do not converge due to very large changes between the iterations.
In one of our previous works, Ref.~\citenum{Pokhilko:algs:2022}, 
 we introduced a heating-and-cooling approach
allowing us to converge such iterations even for very problematic strongly correlated systems. 
Both the original approach and its modification can be viewed as a nonlinear homotopy. 
The heating-and-cooling approach relies on an large increase of temperature, 
which brings the solution to the weakly correlated regime 
close to the exact Hartree--Fock high-temperature limit\cite{Fetter:Walecka:2012,Rusakov14,Welden16,Pokhilko:tpdm:2021}. 
A continuous gradual cooling slowly increases the correlation strength. 
Since the reduction of temperature is gradual, 
the quality of the initial guess for every temperature point is very good. 
Ref.\onlinecite{Iskakov:heating:cooling:2023} further improved the quality of the initial guess 
by performing spine interpolation.  

The main drawback of the heating-and-cooling approach is the lack of control to which solution it converges. 
In the present paper, we fix the issue through the linear homotopy, which we define in our case as
\begin{gather}
H_{\lambda}^{\phi}(G) =  (1-\lambda)\left(-G^{-1} + G_0^{-1} - \Sigma_{\infty}[G]\right) + \nonumber \\ \lambda e^{i\phi}\left(-G^{-1} + G_0^{-1} - \Sigma_{\infty}[G] - \Sigma_{dyn}[G]\right).
\end{gather}
Equation $H_{t}^{\phi}(G) = 0$ is equivalent to  
\begin{gather}
-G^{-1} + G_0^{-1} - \Sigma_{\infty}[G] - \frac{\lambda e^{i\phi}}{(1-\lambda + \lambda e^{i\phi})}\Sigma_{dyn}[G] = 0.
\end{gather}
The constructed parametrization smoothly connects the static self-energy with the full self-energy containing both static and dynamic parts, which allows us to increase the correlation strength gradually. 
At every value of $\lambda$, the restriction on the magnitude of the self-energy acts as a trust region. 

Within this trust region, a linear extrapolation of self-energies through CDIIS is well justified. 
At the same time, homotopy continuation maintains the connection to the Hartree--Fock solution.  
Empirically, we found that only the values of $\phi$ close to zero lead to converging iterations. 
This makes a physical sense since $\phi \neq 0$ can lead to severe violations of causality. 
This essentially restricts the homotopy only to the case of a real homotopy ($\phi = 0$).  

In this manuscript, unless explicitly mentioned, all the calculations used $\phi = 0$. 
The real homotopy has a disadvantage if there is a point where the Jacobian does not have a full rank, 
where a smooth homotopy is not guaranteed to exist.  
In our case, we detected such cases, and they corresponds to developing instabilities. 
However, continuing iterations after such instabilities has allowed us to locate more solutions.  

Starting from a selected guess, we gradually change $\lambda$ converging the Dyson equation (and the chemical potential) 
at every value of $\lambda$. 
For every new value of $\lambda$, we use $\Sigma, G, \mu$ from the previous value as an initial guess. 
We found that if the changes of $G$ and $\Sigma$ with respect to $\lambda$ are relatively small, 
a simple spline extrapolation of $\Sigma, G, \mu$ based on a few previous points can generate a better initial guess, 
which we also use in such cases.

\section{Numerical results}\label{sec:numerical_results}
\subsection{Computational details}\label{sec:comp_details}
We used PySCF 2.2.1\cite{PYSCF,note:pyscf_version} to evaluate one-electron and two-electron integrals and to converge the zero temperature UHF.  
For the Green's function calculations, we used \texttt{Green} package\cite{Green:2025}.  
We used \emph{gth-dzvp-molopt-sr} basis set\cite{GTHBasis}  for Ni, 
\emph{6-311G*} for O\cite{Krishnan:6311Gss:80}, 
and  \emph{gth-dzvp-molopt-sr-q14} for Nd (with g-functions removed due to numerical instabilities in Rys quadrature used for two-electron integral evaluation) taken from the CP2K basis set library. We used \emph{gth-pbe} pseudopotential\cite{GTHPseudo}.  
In all calculations, we used resolution-of-identity approximation (RI) with even-tempered auxiliary basis sets 
generated by PySCF with beta=3.0.  
We applied a diffuse cutoff removing all primitives with exponents smaller or equal to 0.1, 
which is, customarily necessary to avoid linear dependencies in solids.  
We used the $2\times 2\times 2$ Monkhorst--Pack k-point grids for the Brillouin-zone sampling\cite{Monkhorst:Pack:k-grid:1976} for NdNiO$_2$. 
We used the Ewald method\cite{EwaldProbeCharge,CoulombSingular} 
for the finite-size corrections of the Hartree--Fock exchange as implemented in PySCF.  
While the selected k-point grid is small for obtaining definitive quantitative estimates, 
it is sufficient for investigating the convergence toward multiple possible solutions.  
We used the single formula-unit cell (see the atom coordinates and lattice vectors in SI Section~1) obtained as an average of two similar structures from Ref.\cite{Rosseinsky:NdNiO2:structure:2003} 
deposited at the ICSD database with collection codes 98585 and 98586.   
For the Green's function calculations, we used intermediate representation grids\cite{Yoshimi:IR:2017} with $\Lambda = 10^5$.  
We used the inverse temperature $\beta= 1000$~a.u.$^{-1}$ in all the finite-temperature calculations.  
First, for NdNiO$_2$, we converged several PBE\cite{Perdew:96:PBE} solutions.  
To capture them, we requested a different number of $\alpha$ and $\beta$ electrons with and without the ``canonical constraint'' PySCF add-on fixing the number of electrons for every k-point and leading to uniform solutions.  
Then, we used the converged PBE solutions as initial guesses to converge the zero-temperature UHF.  
Subsequently, we used the zero-temperature UHF solutions as initial guesses to converge the finite-temperature UHF. 
We further used the finite-temperature UHF solutions as starting guesses   
for the scGW calculations directly and through the homotopy.   
To converge all the Green's function calculations, 
we employed damping and the frequency-dependent CDIIS algorithm\cite{Pokhilko:algs:2022} with sufficiently tight convergence thresholds to ensure that the reported energies are fully converged.  

To plot NDOs, 
we utilized the PySCF's printer producing orbitals in the Molden format, which we subsequently 
visualized with Gabedit (2.5.1 version)\cite{Gabedit:2011} with isosurface value 0.050 
and rendered with POV-Ray (3.7.0.10 version unofficially ported to Debian)\cite{povray}.

\subsection{Solution search, continuation, and convergence}
To track and compare solutions, we obtain the one-particle density matrix $\gamma$ directly from the Green's function
\begin{gather}
\gamma = - G(\tau=0^{-}). 
\end{gather}
Then we transform the density matrices into Löwdin's orthogonalized SAO basis and evaluate 
\begin{gather}
\sigma_{max} = \max_i \sigma_i (\gamma^{\mathbf{I}, \lambda_{\mathbf{I}}}_{SAO} - \gamma^{\mathbf{J}, \lambda_{\mathbf{J}}}_{SAO}),
\end{gather}
where $\sigma_{i} (M)$ is the $i$-th singular value of a matrix $M$, $\gamma^{\mathbf{I}, \lambda_{\mathbf{I}}}_{SAO}$ 
is the density matrix of \textbf{Solution I} at $\lambda = \lambda_{\mathbf{I}}$ in SAO. 
The largest singular value defines a \emph{spectral norm} of a matrix. 
Due to a variational property of the singular value decomposition, 
the largest singular value gives the largest value of $\braket{\phi_p | (\gamma^{\mathbf{I}, \lambda_{\mathbf{I}}}_{SAO} - \gamma^{\mathbf{J}, \lambda_{\mathbf{J}}}_{SAO}) | \phi_q }$ among all orthonormal orbitals formed from SAO as linear combinations.  
Therefore, it is a convenient measure of a difference between density matrices, 
which also has been actively used for other types of density matrices\cite{Luzanov:TDM-1:76,Luzanov:TDM-2:79,Martin:NTO:03,Luzanov:DMRev:12,Dreuw:ESSAImpl:14,Dreuw:ESSAImpl-2:14,Nanda:NTO:17,Wojtek:ImagEx:18,Krylov:Libwfa:18,Dreuw:NTOfeature:2019,Pavel:SOCNTOs:2019,Nanda:RIXSNTO:20,Wergifosse:respNTO:2020,Nanda:NTO:hyperpolarizability:2021}. 
Since the eigenvalues of the one-particle density matrix are within $[0,1]$ interval, 
any diagonal matrix element is bound by this interval $ 0 \leq \braket{\phi_i | \gamma | \phi_i} \leq 1$. 
Therefore, $0 \leq |\braket{\phi_i | (\gamma^{\mathbf{I}, \lambda_{\mathbf{I}}}_{SAO} - \gamma^{\mathbf{J}, \lambda_{\mathbf{J}}}_{SAO}) | \phi_i }| \leq 1$. 
Hence, the eigenvalues of $\gamma^{\mathbf{I}, \lambda_{\mathbf{I}}}_{SAO} - \gamma^{\mathbf{J}, \lambda_{\mathbf{J}}}_{SAO}$ lay within $[-1,1]$. 
Using the fact that the density matrix is normal,  
the $\sigma_{max}$ value is also within $[0,1]$.  
Our consideration is up to a phase equivalent to a 
natural attachment/detachment analysis of the density matrix differences\cite{HeadGordon:att_det:95}.  
\onecolumngrid

\begin{figure}[t]
  \includegraphics[width=7cm]{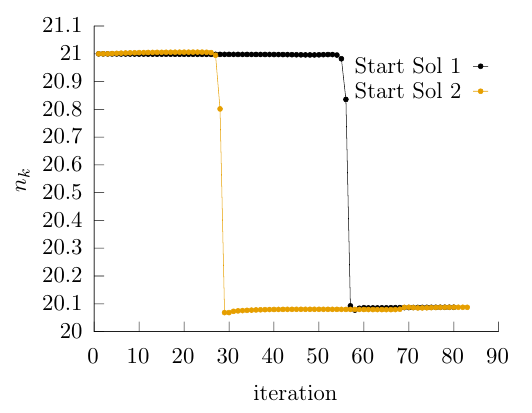}
  \includegraphics[width=7cm]{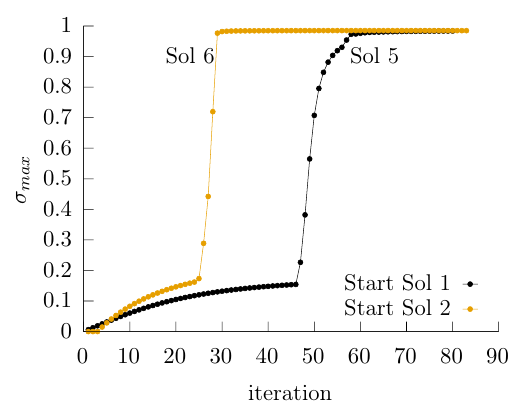}
  \centering
\caption{Flipping of the solution character of the direct scGW iterations with simple damping with $\alpha = 0.1$. 
Left: Occupation numbers at k = (0.424097, 0, 0.506847) and k = (0, 0.424097, 0.506847) are shown 
for the \textbf{Solution 1} and \textbf{2}, respectively.  
Right: The maximum singular values of the density matrix difference between the initial UHF guess and the density matrix at a given iteration. 
\protect\label{fig:NdNiO2_naive}
}
\end{figure}
\twocolumngrid

Following the procedure in Section~\ref{sec:comp_details}, 
we obtained the following finite-temperature UHF solutions: 
\begin{itemize}
\item \textbf{Solution 1} with $S_z = 1$ per unit cell with the same number of electrons for different k-points
\item \textbf{Solution 2} with $S_z = 1$ per unit cell with different numbers of electrons for different k-points
\item \textbf{Solution 3} with $S_z = 2$ per unit cell with the same number of electrons for different k-points 
\item \textbf{Solution 4} with $S_z = 2$ per unit cell with different numbers of electrons for different k-points
\end{itemize}
Table~S1 in the SI shows the occupation numbers for every k-point for all these UHF solutions. 

\begin{figure}[H]
  \includegraphics[width=7cm]{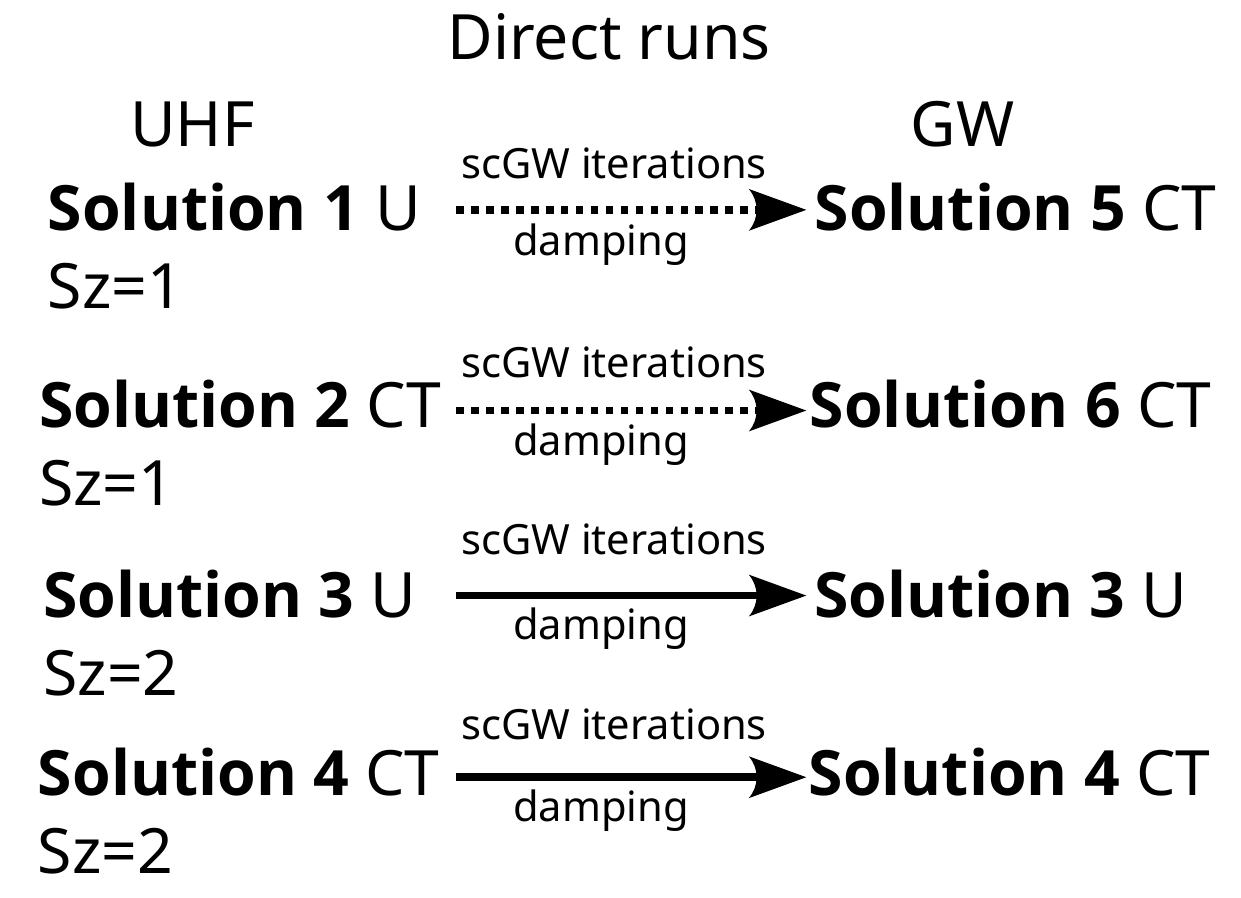}
\caption{Behavior of the damping scGW iterations. The dashed arrow represents a switch from one solution to the other; the solid arrow represents iterations preserving the character of solutions.}
\protect\label{fig:direct_runs}
\end{figure}

\subsubsection{Direct scGW calculations starting from an UHF guess}

Starting from the prepared UHF solutions, we attempted the direct damped scGW iterations. However, the results of these iterations do not necessarily correspond to the same physical solutions as the starting points (Fig.~\ref{fig:direct_runs}). 
The direct damped scGW iterations starting from UHF \textbf{Solutions 3} and \textbf{4} converge 
to the scGW solutions with the same respective character.  

The direct damped scGW iterations starting from UHF \textbf{Solutions 1} and \textbf{2} converge to the scGW solutions  
with characters different from the starting UHF points. These scGW solutions move $\beta$ electrons between k-points (Fig~\ref{fig:NdNiO2_naive}). We denote these scGW solutions as \textbf{Solutions 5} and \textbf{6}.   
\onecolumngrid

\begin{figure}[!h]
  \includegraphics[width=7cm]{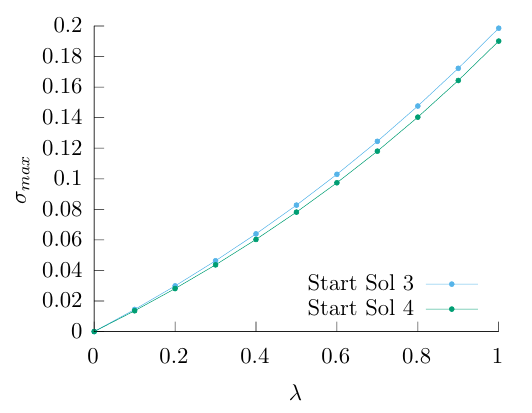}
  \includegraphics[width=7cm]{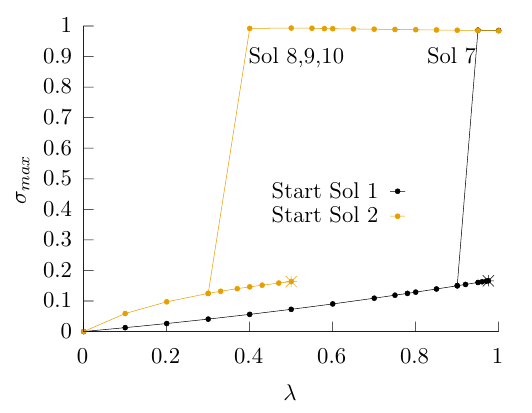}
  \centering
\caption{Singular values of the difference between the density matrices of the initial UHF and the fully self-consistent solution at different values of the homotopic parameter $\lambda$.  
\protect\label{fig:NdNiO2_4_homotopy}
}
\end{figure}
\begin{figure}[H]
  \includegraphics[width=12cm]{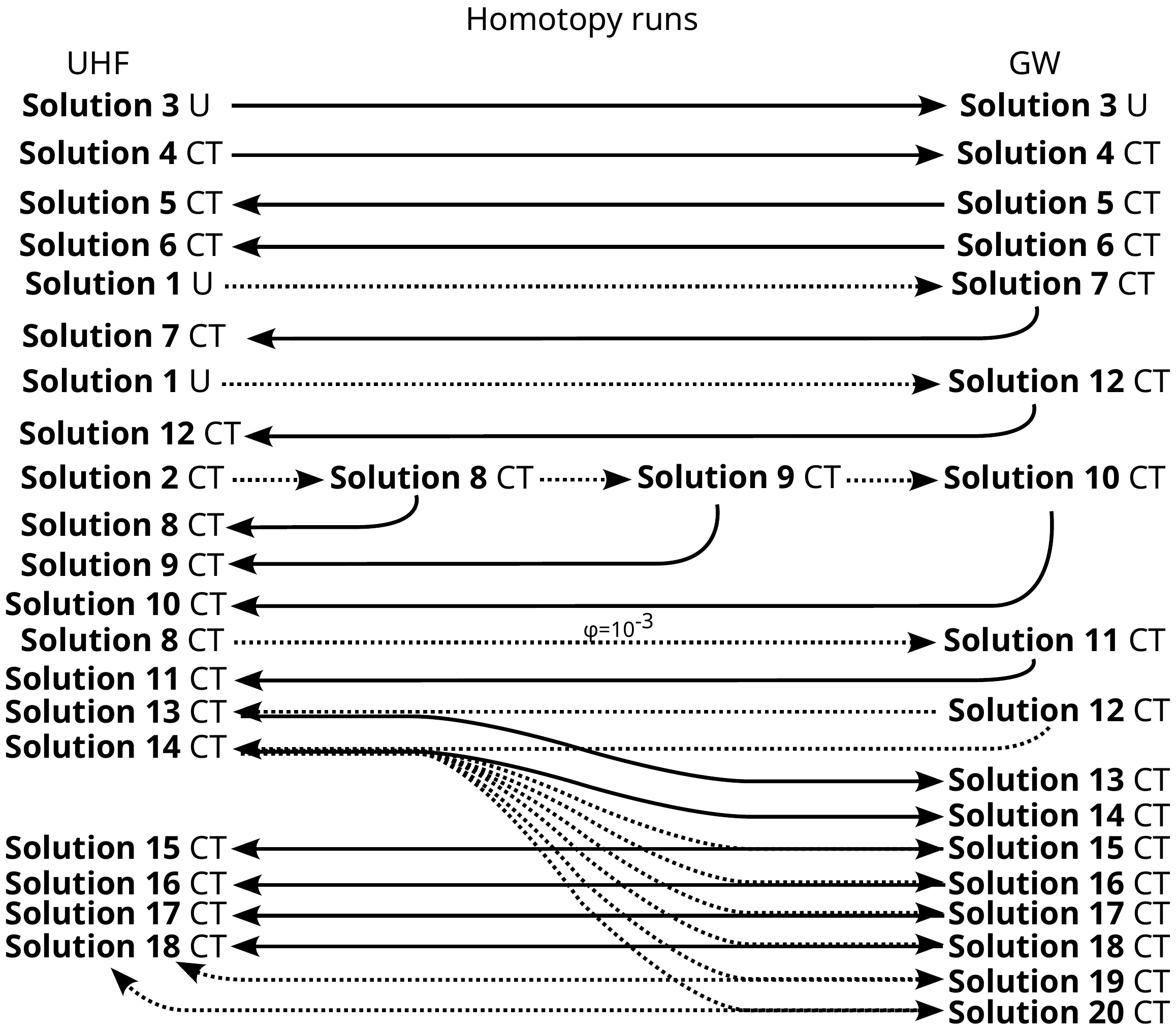}
  \centering
\caption{Behavior of the homotopy continuation. Dashed arrows represent continuation attempts that switched from one solution to the other; the solid arrows represent homotopy continuations preserving the character of solutions.}
\protect\label{fig:homotopy_runs}
\end{figure}
\twocolumngrid

\subsubsection{Homotopy}

Fig.~\ref{fig:homotopy_runs} shows the overall structure of the homotopy continuations as well as the continuation attempts. We provide the details of these attempts in the text below. 

The homotopic continuations of UHF \textbf{Solutions 3} and \textbf{4} to the corresponding scGW solutions are smooth (see Fig.~\ref{fig:NdNiO2_4_homotopy}), but take a larger number of iterations than the directly running scGW. 
The larger number of iterations is expected because in the homotopy method the step sizes are restricted at a fixed value of $\lambda$.  
We verified that both the direct runs and the homotopic runs resulted in the convergence to the same solutions. 

The homotopic continuations of the UHF \textbf{Solutions 1} and \textbf{2} are more sensitive to the homotopic step 
requiring monitoring and possible adjustments of the step size in the problematic regions of the homotopy.  
However, even with very small step sizes for such solutions there are such values of $\lambda$ at which we could not converge solutions. We show such points with the star symbol on the graphs.  
It is likely that around these points the solutions become too unstable and  the iterations cannot be continued.  
Damping iterations near such points led to switching from \textbf{Solutions 1} and \textbf{2} to \textbf{Solutions 7} and \textbf{8}, respectively. 
A subsequent continuation of \textbf{Solution 8} led to consequent switching first to \textbf{Solution 9}, 
then to  \textbf{Solution 10}. 
One of the continuation attempts of \textbf{Solution 1}
switched to \textbf{Solution 12}.  
\onecolumngrid

\begin{figure}[!h]
  \includegraphics[width=7cm]{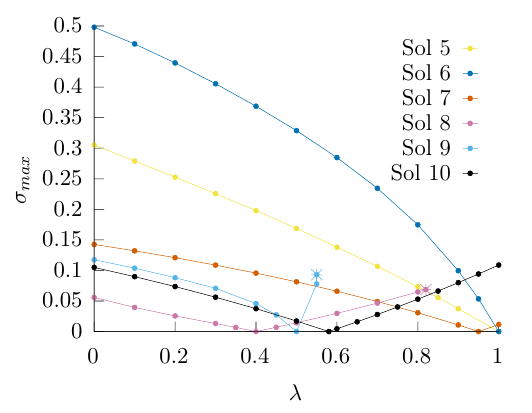}
  \includegraphics[width=7cm]{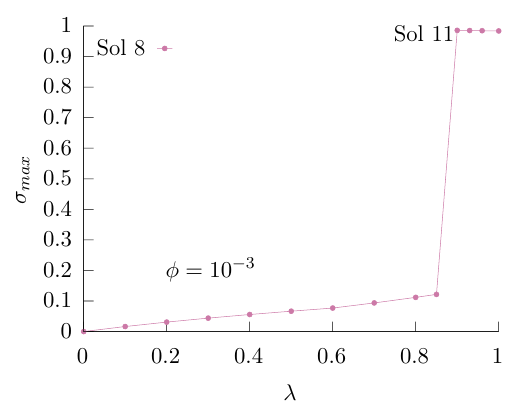}
  \includegraphics[width=7cm]{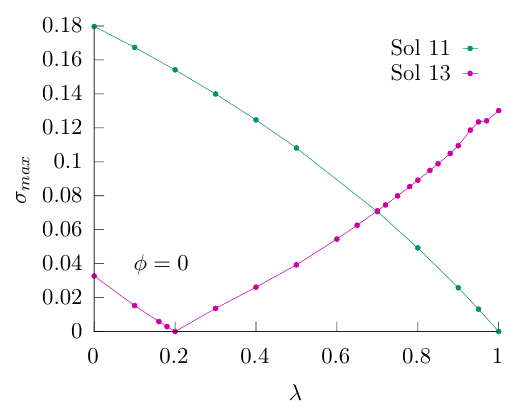}
  \includegraphics[width=7cm]{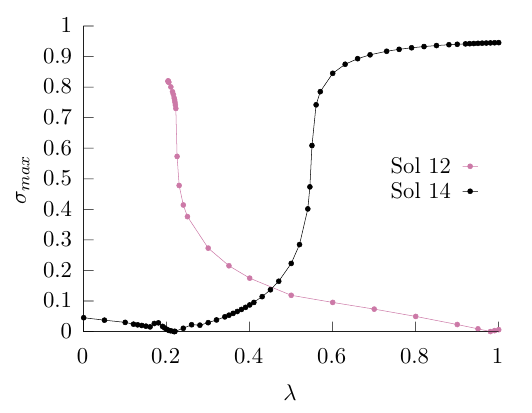}
  \centering
\caption{Top left: Homotopic continuation from the ``switched'' solutions ($\phi = 0$). 
         Top right: Continuation from the UHF \textbf{Solution 8} along the complex path with $\phi = 10^{-3}$. 
         Bottom left: Continuation  along the real path ($\phi = 0$) from the scGW  \textbf{Solution 11} and \textbf{13}.
         Bottom right: Continuation of \textbf{Solution 12} and \textbf{14}, ($\phi = 0$).
\protect\label{fig:NdNiO2_cont_homotopy}
}
\end{figure}
\twocolumngrid

We investigated these solutions further (Fig.~\ref{fig:NdNiO2_cont_homotopy}, Top left) 
and homotopically continued them toward the increase and decrease of $\lambda$. 
Most of the solutions found  have continuity to the UHF limits.

It was possible to continue \textbf{Solutions 5--7, 10} to the corresponding scGW limits. 
At certain values of $\lambda$ it was impossible to converge \textbf{Solutions 8, 9}, 
which likely became unstable. 
Starting from the UHF limit, we tried to continue \textbf{Solution 8} along the complex path, 
which led to switching to a new solution, which we denote as \textbf{Solution 11} (Fig.~\ref{fig:NdNiO2_cont_homotopy}, Top right). 
We obtained a smooth continuation from the scGW limit of \textbf{Solution 11} to the UHF limit along the real homotopy. 

We show the continuation of \textbf{Solution 12} in Fig.~\ref{fig:NdNiO2_cont_homotopy}. 
This is the only solution that cannot be fully continued toward the UHF limit.  
\textbf{Solution 12} changes its character along continuation and around $\lambda = 0.2$ becomes very sensitive to the step size, switches to other solutions (\textbf{Solutions 13} and \textbf{14}), and does not converge.  
We continued \textbf{Solutions 13} and \textbf{14} toward the UHF and GW limits along the homotopies. 
\textbf{Solution 14} smoothly changes its physical character along homotopy. 
\onecolumngrid
\begin{figure}[!h]
  \includegraphics[width=7cm]{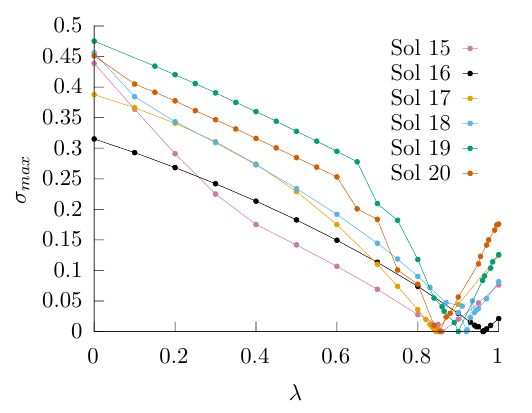}
  \centering
\caption{Homotopy continuation of \textbf{Solutions 15--20}. 
\protect\label{fig:NdNiO2_homotopy_15}
}
\end{figure}
\twocolumngrid
Initially we took too large $\Delta \lambda$ steps for the homotopy of \textbf{Solution 14}, 
which switched to \textbf{Solutions 15, 16, 17, 18, 19,} and \textbf{20}. 
\textbf{Solutions 18, 19, 20} have the same UHF limit, but different scGW limits, 
which we checked through singular values of density matrix differences.

\subsection{Characters of the solutions and their physical meaning}
We evaluated $\sigma_{max}$ for all of the solution pairs 
(Tables~S13 and S14 in SI), 
which ensured that all of the found solutions are distinct (except UHF \textbf{Solutions 18, 19, 20}).   
We show total energies in a.u. of the obtained solutions in Table~\ref{tbl:tot_energies}.  
For almost all distinct pairs of the UHF solutions, the values of $\sigma_{max}$ 
approach 1 meaning that solutions differ from each other by a transfer of at 
least one electron from one orthogonal orbital to the other. 
The only exception from this observation are \textbf{Solutions 5} and \textbf{9} with $\sigma_{max} = 0.53$. 
At the scGW limit these observations are qualitatively the same with 
the values of $\sigma_{max}$ are only slightly less than 1.  
\textbf{Solutions 18--20} at the scGW limit have non-zero $\sigma_{max}$ of 0.28 and 0.52.  

For the cases when both UHF and scGW limits can be found, we evaluated $\sigma_{max}$ between the UHF and scGW solutions. 
This value can quantify how incorporation of correlation by means of scGW changes the density matrix. 
For \textbf{Solutions 3, 4, 5, 7, 10, 11, 13, 16}, $\sigma_{max}$(UHF,scGW) is within 0.15--0.34, 
corresponding to a rather modest change within a weak/intermediate correlation.   
For \textbf{Solutions 6, 15, 17, 18--20},  $\sigma_{max}$(UHF,scGW) is within 0.46--0.58, 
corresponding to a transfer of a half-electron, which is already a substantial correlation. 
Finally, for \textbf{Solution 14}, $\sigma_{max}$(UHF,scGW) is 0.96, 
meaning that correlation has changed its character completely. 
As clear from Fig.~\ref{fig:NdNiO2_cont_homotopy} right bottom, 
the change of character of \textbf{Solution 14} with respect to the homotopic parameter happened smoothly.

\begin{table*} [tbh!]
  \caption{Total energies in a.u. of the obtained solutions
}
\protect\label{tbl:tot_energies}
\begin{tabular}{l|cc||l|cc}
\hline
\hline
\# sol.   & UHF  & scGW     & \# sol.              & UHF  & scGW  \\    
\hline
 1 & -255.018429 & ---          & 11& -255.055503 & -256.295824 \\
 2 & -255.017912 & ---          & 12&  ---        & -256.282207 \\
 3 & -255.087209 & -256.310540  & 13& -255.057296 & -256.291025 \\
 4 & -255.088608 & -256.318665  & 14& -255.042446 & -256.287295 \\
 5 & -255.047114 & -256.276711  & 15& -255.073949 & -256.295991 \\
 6 & -255.055930 & -256.278172  & 16& -255.069063 & -256.288928 \\
 7 & -255.048319 & -256.277648  & 17& -255.079523 & -256.297824 \\
 8 & -255.027806 & ---          & 18& -255.069040 & -256.295934 \\
 9 & -255.041090 & ---          & 19& -255.069040 & -256.295991 \\
 10& -255.084701 & -256.315907  & 20& -255.069038 & -256.295989  \\ 
\hline
\hline
\end{tabular}
\end{table*}
While we could not find the scGW solutions corresponding to the high-energy UHF solutions, 
for most of the low-energy solutions such a correspondence is established.  
Table~\ref{tbl:tot_energies} shows energies of all the found solutions.  
It is clear that there are multiple nearly degenerate solutions, which are 
sometimes different only by a few K (1 K $= 3.166 811 \cdot 10^{-6}$~a.u. by NIST \cite{NIST:unit_converter}).

When applied to strongly correlated problems, the Hartree--Fock theory gives multiple solutions. 
While the Hartree--Fock theory cannot capture the strong electron correlation within the same solution, this set of solutions often can reveal the essential components of the strong electron correlation at a qualitative level\cite{Fukutome:UHF:81}. 
Weakly correlated methods often preserve the structure of the Hartree--Fock solutions and can even be used to reconstruct the structure of strong electron correlation even at a quantitative level, such as within the broken-symmetry approach\cite{Yamaguchi:UHFS:1979,noodleman:BS:81,Yamaguchi:BS:formulation:1986,Pokhilko:local_correlators:2021,Pokhilko:BS-GW:solids:2022,Pokhilko:Neel_T:2022,Pokhilko:THC-G3W2:2025}. We located multiple solutions with a charge-transfer character within a small energy window meaning that the excitation energy inducing the charge transfer is small.

Tables~S1 and S12 in SI show the number of electrons for every k-point. 
For all the corresponding solutions, going from UHF to correlated scGW does not alter k-point occupations significantly.  
However, different solutions may have very different k-point occupations. 
Only \textbf{Solutions 1} and \textbf{3} have equal numbers of $\alpha$ and $\beta$ electrons for every k-point 
(we call them uniform solutions).  
All other solutions have unequal distributions of electrons across k-points either for $\alpha$ spin, 
$\beta$ spin, or both, giving rise to charge density waves (CDW) or spin-charge density waves (SCDW).  
In comparison to the uniform solutions, \textbf{Solution 2} transfers one of the $\beta$ electrons to $\Gamma$ point, 
while \textbf{Solution 4} transfers one of the $\alpha$ electrons from a different k-point to $\Gamma$ point. 
By the number of $\alpha$ and $\beta$ electrons for each k-point, 
\textbf{Solutions 2, 5, 8}, and \textbf{9} cannot be distinguished from each other (but they differ by local occupations within k-points, see below). 
\textbf{Solution 10} differs from \textbf{2,5,8,9} by the k-point of the ``hole'' 
from which a $\beta$ electron is transferred to the $\Gamma$ point. 
\textbf{Solution 6} transfers two $\beta$ electrons. 
\textbf{Solution 11} transfers 3 $\beta$ electrons. 
Other solutions flip spins of one or several electrons. 
\textbf{Solution 7} transfers one $\beta$ electron from one k-point to the $\Gamma$ point 
and another $\beta$ electron to a different k-point flipping the spin. 
\textbf{Solution 13} transfers 3 $\beta$ electrons to different k-points flipping the spin of two of them. 
\textbf{Solutions 13, 14, 15}, and \textbf{17} have the same distributions of electrons for k-points. 
\textbf{Solution 16} transfers 4 $\beta$ electrons flipping the spin of two of them. 
\textbf{Solution 18} transfers 3 $\beta$ electrons flipping the spin of only one of them. 
\textbf{Solution 12} (found only for scGW)  transfers 3 $\beta$ electrons flipping the spin of two of them. 
GW \textbf{Solutions 18, 19, 20} have identical distributions of electrons across k-points. 

\subsubsection{Implications of charge transfer}
As clear from the charge-transfer character (CDW) of the obtained solutions, 
transfer of one or even several electrons does not change the energy much. 
This means that these configurations are energetically accessible, 
which can enable a substantial conductivity.  
Experimentally measured electric resistivity of NdNiO$_2$ is 
around 2~m$\Omega \cdot$cm \cite{Hwang:NdNiO2:conductivity:2020},  
which is several times lower than the resistivity of polycrystalline graphite\cite{Wilson:graphite:1953}.   
Thus, the found solutions give a qualitative prediction consistent with this experiment.  
Experimentally, CDW character of the ground state was attributed 
based on the analysis of RIXS experiments\cite{Zhou:CDW:NdNiO2:2022}, which, however, have been debated in the subsequent experiments\cite{Shen:noCDW:NdNiO2:2024}.  
Theoretically, a formation of CDW was proposed in Ref.~\cite{Deng:NdNiO2:2024} based on DFT calculations. 
We would like to clarify that the CDW in our work differs qualitatively from Ref.~\cite{Deng:NdNiO2:2024}, 
where the CDW character was induced by the lattice distortions of the undoped compound. 
In our work, the CDW is the character of the solution itself occurring due to electronic instabilities  
at the high-symmetry geometry. 
Based on the comparison of the charge-transfer p--d gap with cuprates within DFT,  
NdNiO$_2$ was classified as a weak Mott insulator\cite{Sun:NdNiO2:2021}.   
Advanced calculations based, however, on a DFT-derived 3-orbital model   
also ruled out a possibility of CDW ordering\cite{Stepanov:NdNiO2:2024}.  
Our results do not support this classification.  
The energy difference between some of the solutions obtained, corresponding to an electron transfer, is very small. 
For example, the difference between scGW \textbf{Solution 5} and \textbf{6} is just 0.04~eV. 
We also observe additional differences from cuprates based on participating orbitals that we discuss in the next section. 

Very recently, the CDW was attributed based on variational cluster approximation within a DFT-derived two-orbital model in Ref.\cite{Oles:NdNiO2:VCA:2025}. In our calculations we do not reduce the number of orbitals, which has allowed us to gain the new insights of the charge-transfer physics of this material, as we show in the sections below. 

\subsection{Participating orbitals}
There are multiple and often contradictory reports about participating orbitals 
including the role of f-, and d-orbitals for both Nd and Ni; s-orbitals for Nd, and p-orbitals for O. 
Partly, these contradictions are caused by an extensive use of plane-wave basis\cite{Deng:NdNiO2:2024,Millis:NdNiO2:2023,Millis:NdNiO2:2020}, 
where a projection is needed to determine the role of an atomic orbital with a certain angular momentum. 
We employ a basis of atomic orbitals that use Gaussian orbitals.
This means that can provide information about orbitals participating in the unit cell without ambiguities caused by projection and Wannierization techniques. 
Below we provide a comprehensive analysis of our solutions elucidating the participation of different orbitals.  

\begin{table*} [tbh!]
  \caption{Frobenious norms of the differences between $P_{Nd_{4f}}\gamma_{SAO}^{\mathbf{I}}P_{Nd_{4f}}$ and $P_{Nd_{4f}}\gamma_{SAO}^{\mathbf{1}}P_{Nd_{4f}}$ for the UHF solutions, where $P_{Nd_{4f}}$ is the projector onto the $4f$ SAO on Nd.  Note that the table shows the difference for every UHF solution I with respect to the first solution.
}
\protect\label{tbl:f_Fnorm}
\begin{tabular}{l|c|l|c|l|c|l|c}
\hline
\# sol.  & Frob. norm & \# sol. & Frob. norm & \# sol. & Frob. norm &  \# sol. & Frob. norm \\
\hline
\hline
1  & 0.000  &  6  & 0.042   &   11 & 0.021  &  17 & 0.047 \\
2  & 0.022  &  7  & 0.033   &   13 & 0.017  &  18 & 0.025 \\
3  & 0.066  &  8  & 0.016   &   14 & 0.045  &  19 & 0.025 \\
4  & 0.064  &  9  & 0.035   &   15 & 0.042  &  20 & 0.025 \\
5  & 0.039  &  10 & 0.061   &   16 & 0.061  &     &       \\   
\hline
\hline
\end{tabular}
\end{table*}
\subsubsection{Role of f-orbitals}
Although the f-orbitals were included in pseudopotentials in a number of publications\cite{Alavi:NdNiO2:2022,Deng:NdNiO2:2024,Millis:NdNiO2:2023,Millis:NdNiO2:2020,Yang:NdNiO2:2020}, 
a few publications raised a question about a potential role of f-orbitals\cite{Pickett:4f:NdNiO2:2020,Sun:NdNiO2:2021,Bandyopadhyay:NdNiO2:2020,Oles:NdNiO2:f-orbs:2023}.  We used the GTH pseudopotential, which keeps the f-orbitals in the valence region. 
To check how different the solutions are due to the participation of f-orbitals, 
we transformed the density matrices to SAO, took the blocks corresponding to $4f$ orbitals for Nd at every k-point, 
and evaluated the Frobenius norm of their differences between different solutions (Table~\ref{tbl:f_Fnorm}).  
The Frobenius norm allows us to detect even small changes distributed across different nearly degenerate f-orbitals. 
All the norms are very small, meaning that the solutions share almost the same occupations of $4f$ orbitals. 
While adjustments of $4f$ orbitals may play some quantitative role of small interactions, 
from a qualitative perspective, they do not play a significant role for the charge transfer. 
The small participation within UHF and GW is additionally confirmed through the analysis of natural difference orbitals (see below) and Tables S15 and S14 in SI. 

\begin{table*} [tbh!]
  \caption{Frobenious norms of the differences between $P_{Nd_{5s}}\gamma_{SAO}^{\mathbf{I}}P_{Nd_{5s}}$ and $P_{Nd_{5s}}\gamma_{SAO}^{\mathbf{1}}P_{Nd_{5s}}$ for the UHF solutions, where $P_{Nd_{5s}}$ is the projector onto the $5s$ SAO on Nd.  
}
\protect\label{tbl:s_Fnorm}
\begin{tabular}{l|c|l|c|l|c|l|c}
\hline
\# sol.  & Frob. norm & \# sol. & Frob. norm & \# sol. & Frob. norm &  \# sol. & Frob. norm \\
\hline
\hline
1  & 0.000  &  6  & 0.030   &   11 & 0.020  &  17 & 0.021 \\
2  & 0.003  &  7  & 0.019   &   13 & 0.020  &  18 & 0.032 \\
3  & 0.025  &  8  & 0.019   &   14 & 0.020  &  19 & 0.032 \\
4  & 0.025  &  9  & 0.019   &   15 & 0.020  &  20 & 0.032 \\
5  & 0.020  &  10 & 0.023   &   16 & 0.032  &     &       \\   
\hline
\hline
\end{tabular}
\end{table*}
\subsection{Role of s-orbitals on Nd}
Similarly to the investigation of the role of f-orbitals, 
we evaluated the same norms of the differences for the 5s orbitals on Nd (Table~\ref{tbl:s_Fnorm}). 
The resulting norms are very small showing that 5s is not involved qualitatively in the charge transfer.  

\begin{table*} [tbh!]
  \caption{Frobenious norms of $P_{O_{p}}(\gamma_{SAO}^{\mathbf{I}}-\gamma_{SAO}^{\mathbf{1}})P_{O_{p}}$ for the UHF solutions.  
}
\protect\label{tbl:O_Fnorm}
\begin{tabular}{l|c|l|c|l|c|l|c}
\hline
\# sol.  & Frob. norm & \# sol. & Frob. norm & \# sol. & Frob. norm &  \# sol. & Frob. norm \\
\hline
\hline

1 & 0.000  &  6  & 0.328   & 11 & 0.450   &  17 & 0.309  \\
2 & 0.066  &  7  & 0.274   & 13 & 0.295   &  18 & 0.328  \\
3 & 0.496  &  8  & 0.293   & 14 & 0.278   &  19 & 0.329  \\
4 & 0.554  &  9  & 0.358   & 15 & 0.280   &  20 & 0.329  \\
5 & 0.315  &  10 & 0.518   & 16 & 0.335   &     &  \\   
\hline
\hline
\end{tabular}
\end{table*}
\subsubsection{Role of s-orbitals and p-orbitals on O}
We evaluated the same norms of the differences for all the s-orbitals on O, 
which turn out to be less than $0.07$ in magnitude. 
Thus, qualitatively, $s$-orbitals do not contribute much to the charge transfer between solutions. 
We show the norms of the differences for the atomic p-orbitals on O in Table~\ref{tbl:O_Fnorm}. 
The magnitude of their contribution points to a strong hybridization 
of the relevant d-orbitals with p-orbitals on oxygen. 
As seen from the natural difference orbitals below, 
p-orbitals on O indeed actively hybridize with the d$_{x^2-y^2}$ on Ni. 
The s-orbitals on O hybridize with the d-orbitals on Ni as well, but to a much smaller degree. 

\subsubsection{Role of d-orbitals on Nd and Ni}
\begin{figure}[!h]
  \includegraphics[width=7cm]{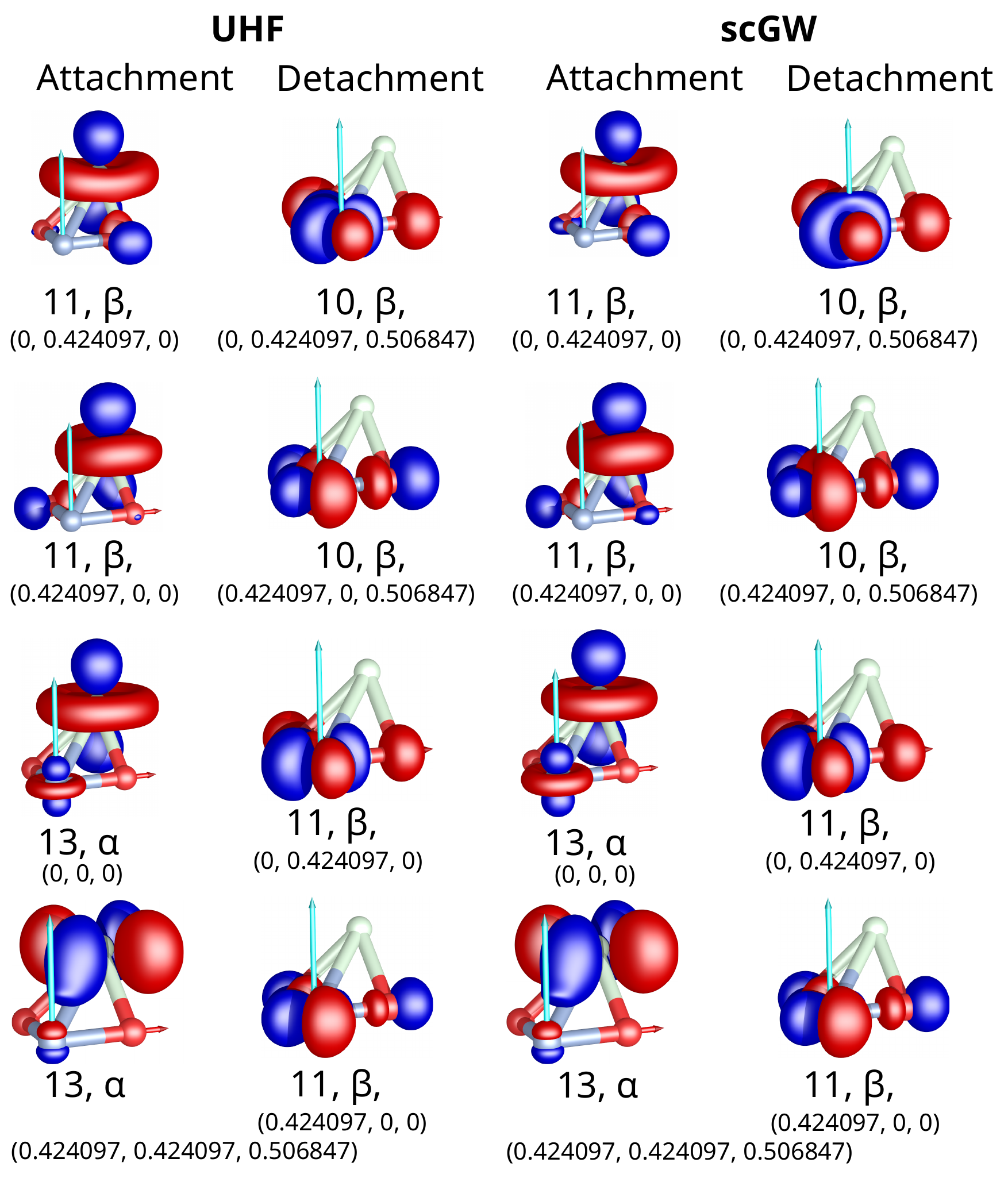}
  \centering
\caption{Leading spin-resolved natural difference orbitals for \textbf{Solutions 11,10} and \textbf{13,11}. Orbital spin, and the corresponding k-points are shown. 
\protect\label{fig:NDO3}
}
\end{figure}

\begin{figure}[!h]
  \includegraphics[width=5cm]{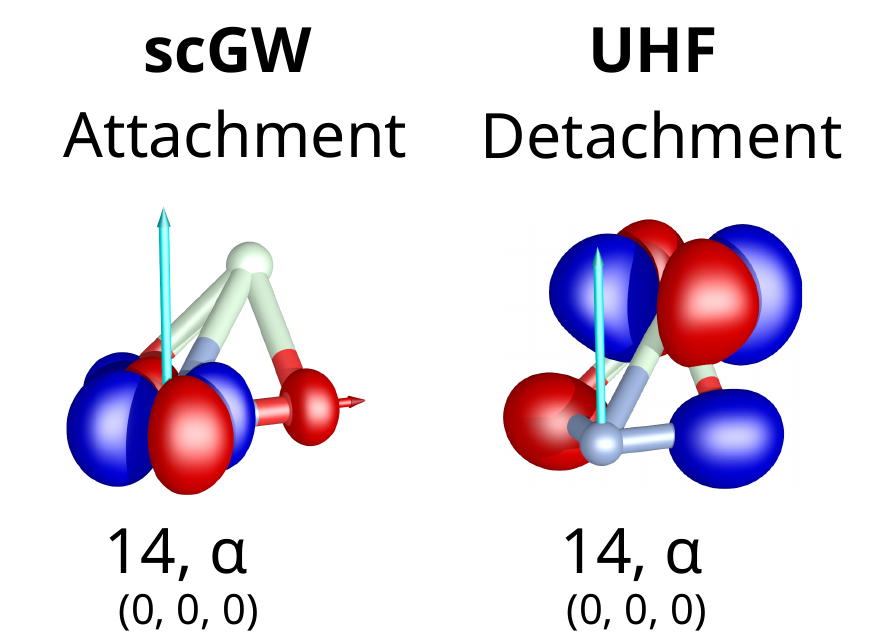}
  \centering
\caption{Leading spin-resolved natural difference orbitals for scGW and UHF limits of \textbf{Solutions 14}.  
\protect\label{fig:NDO4}
}
\end{figure}

\begin{figure}[!h]
  \includegraphics[width=7cm]{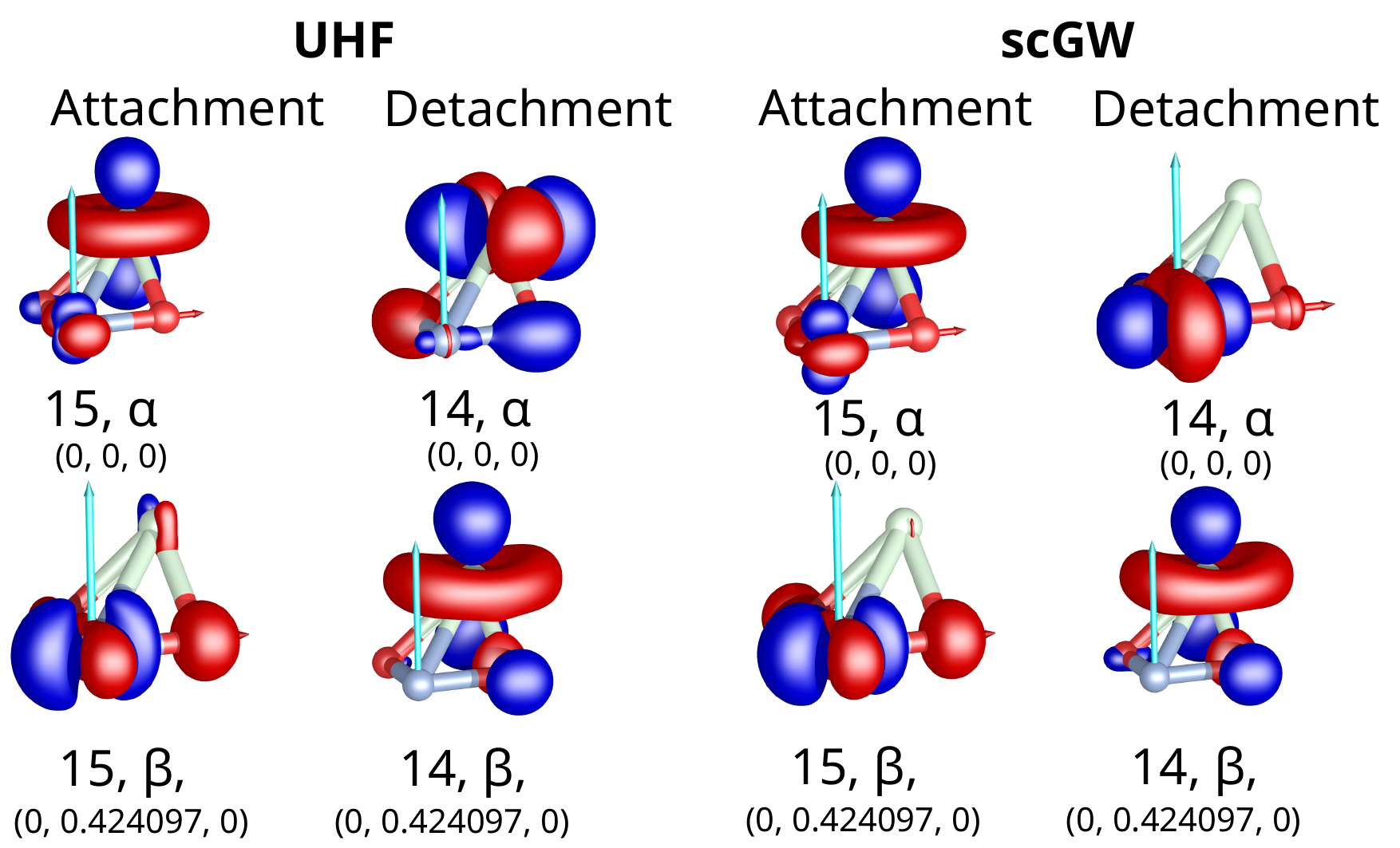}
  \centering
\caption{Leading spin-resolved natural difference orbitals for scGW and UHF limits of \textbf{Solutions 15, 14}.  
\protect\label{fig:NDO5}
}
\end{figure}

\begin{figure}[!h]
  \includegraphics[width=7cm]{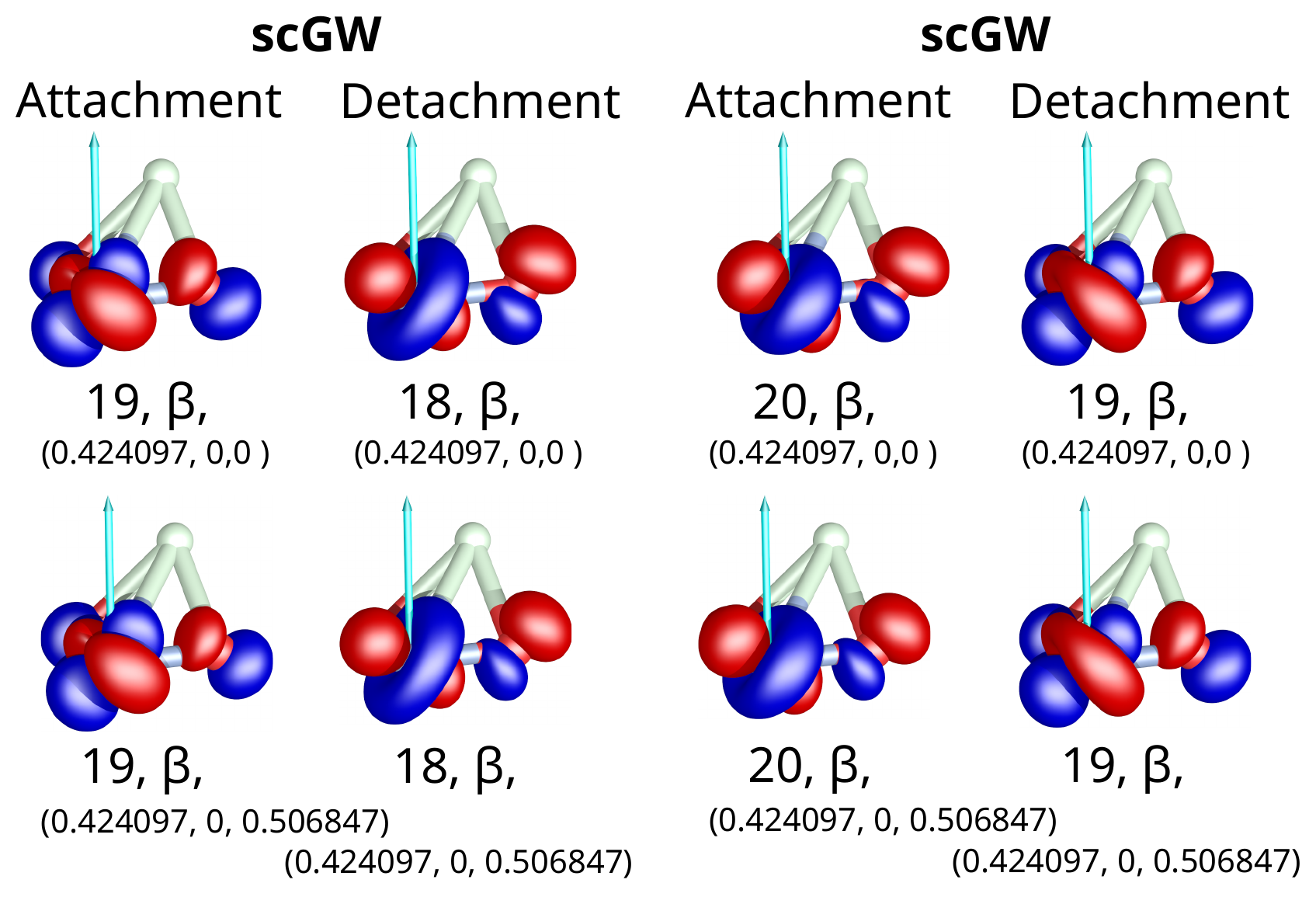}
  \centering
\caption{Som of the leading spin-resolved natural difference orbitals for scGW and UHF limits of \textbf{Solutions 18, 19, 20}.  
\protect\label{fig:NDO6}
}
\end{figure}

. The charge transfer between the d-orbitals is 
the key distinction between the solutions. 
In SI, Tables~S2--S7 and Tables~S8--S11 show the d-orbital Lowdin populations for UHF solutions and scGW solutions, respectively.
While these tables are informative, a clear and concise picture can be gained 
from the natural difference orbitals defined as eigenvectors 
between the difference of density matrices in the k-space.  
This concept extends the analysis done in Ref. \cite{Pokhilko:NO:correlators:2023} to the case of several solutions with very different orbital occupations. We define the NDOs in the k-space as eigenvectors of $\gamma^{\mathbf{I}}_{SAO} - \gamma^{\mathbf{J}}_{SAO}$. Similarly to the definition for molecules\cite{HeadGordon:att_det:95}, we label the NDOs with the positive eigenvalues as ``attachment'' and the NDOs with the negative eigenvalues as ``detachment''. 
\onecolumngrid
\begin{figure}[!h]
  \includegraphics[width=7cm]{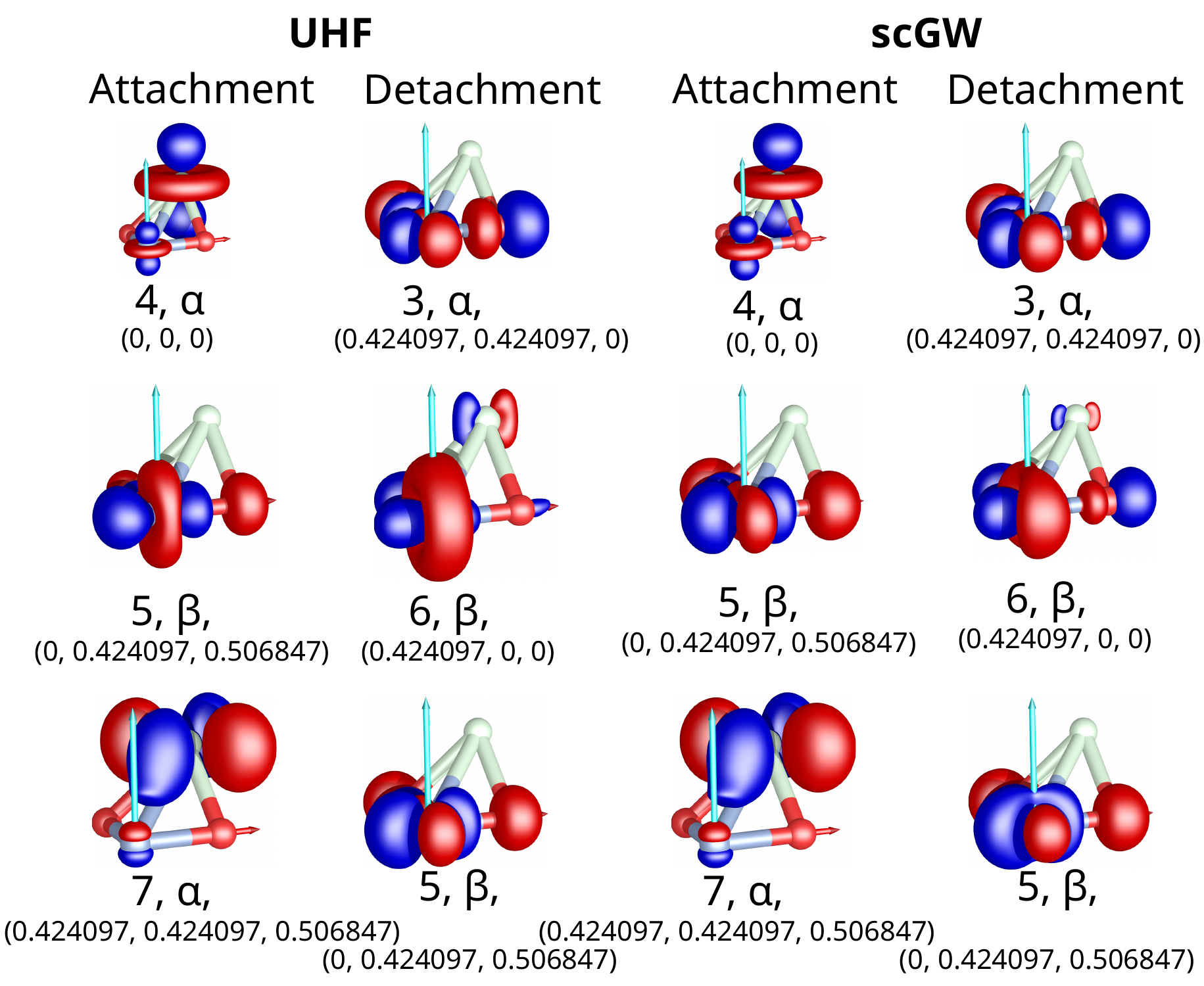} \hfill
  \includegraphics[width=7cm]{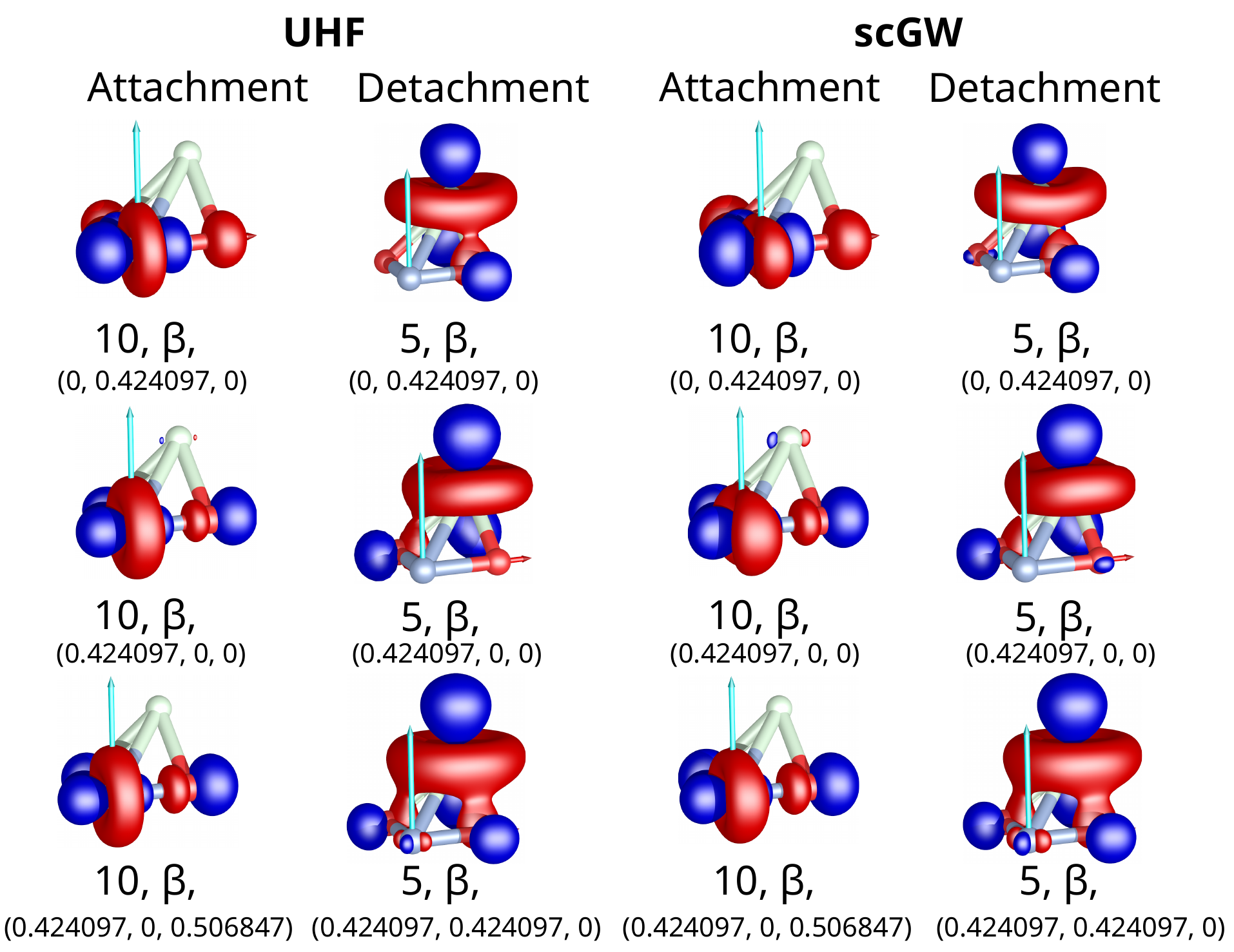}
  \centering
\caption{Leading spin-resolved natural difference orbitals for a few selected pairs of solutions. Solution number, orbital spin, and the corresponding k-points are shown. Multiple NDO pairs contribute to the difference between \textbf{Solutions 10} and \textbf{5}. 
\protect\label{fig:NDO1}
}
\end{figure}
\twocolumngrid
We show such an analysis of the leading NDOs in Fig.~\ref{fig:NDO1}--\ref{fig:NDO6}.  
For all the solutions, the leading atomic coefficients in NDOs are located on 5d$_{z^2}$, 5d$_{xy}$ orbitals on Nd and on 
3d$_{x^2-y^2}$, 3d$_{z^2}$, 3d$_{x^2}$, 3d$_{y^2}$ orbitals on Ni. 
Although the imaginary part of NDOs is not zero, it is very small and is neglected for the simplicity of visualization.  
The solutions have a different structure with respect to orbitals participating in the charge transfer.  

\textbf{Solutions 4} and \textbf{3}, the solutions with the highest number of $\alpha$ electrons, are different only by a location of one $\alpha$ electron. It occupies a 3d$_{x^2-y^2}$ orbital on Ni strongly hybridized with the p-orbitals on O in \textbf{Solution 3} and occupies the predominantly 5d$_{z^2}$ orbital in \textbf{Solution 4}. 
scGW leads to a larger charge delocalization slightly increasing contribution of 3d$_{z^2}$ orbital on Ni in \textbf{Solution 4}. This is consistent with the effect of \emph{screening} introduced by $W$. 
We observed such increases of weights from ligand orbitals before \cite{Pokhilko:NO:correlators:2023}. 
The scGW increases the energy difference (Table~\ref{tbl:tot_e_diff}) stabilizing charge-transfer \textbf{Solution 4} as the lowest solution that we found. 
\begin{table*} [tbh!]
  \caption{Total energy difference (eV) between selected \textbf{I,J} solutions taken as $E_{I} - E_{J}$. 
}
\protect\label{tbl:tot_e_diff}
\begin{tabular}{l|cc||l|cc}
\hline
\hline
Pair   & UHF  & scGW       & Pair    & UHF  & scGW  \\    
\hline
4,3    & -0.04 & -0.22     &  13,11  & -0.05 &  0.13   \\    
5,6    &  0.02 &  0.04     &  15,14  & -0.86 & -0.24   \\
7,5    & -0.03 & -0.03     &  19,18  &  0.00 &  0.00   \\
10,5   & -1.02 & -1.07     &  20,19  &  0.00 &  0.00   \\
11,10  &  0.79 &  0.55     &         &       &         \\
\hline
\hline
\end{tabular}
\end{table*}

Nearly degenerate \textbf{Solutions 5} and \textbf{6} are different only by the location of $\beta$ electron on Ni $d_{x^2-y^2}$ and $d_{x^2}$ orbitals within different k-points. 
Although these solutions are higher in energy, they suggest that the charge transfer between Ni centers is thermodynamically plausible. 
Inclusion of screening through scGW once again increases the contributions from oxygen. 

Nearly degenerate \textbf{Solutions 5} and \textbf{7} are different by a spin-flipping excitation moving a $\beta$ electron from Ni d$_{x^2-y^2}$ orbital to Nd d$_{xy}$ orbital. 

There are multiple leading NDOs for \textbf{Solutions 10} and \textbf{5} for different k-points. 
However, their atomic structure is very similar caused by 
a charge transfer from Nd d$_{z^2}$ orbital hybridized with O to Ni d$_{x^2}$ orbital also hybridized with O.  
scGW does not change the energy difference between these solutions significantly, 
which can be explained by the fact that for both solutions the distribution of electrons by k-points is different from the uniform solution only by a move of one electron between k-points. 

\textbf{Solutions 11} and \textbf{10} are different by the locations of two $\beta$ electrons between the d$_{z^2}$ orbital on Nd and the hybridized d$_{x^2-y^2}$ orbital on Ni. 
\textbf{Solution 11} is much higher in energy, which is consistent with its high charge-transfer character (3 electrons are moved between k-points in comparison with the uniform distribution across k-points). 
scGW reduces the energy difference between \textbf{Solutions 11} and \textbf{10} indicating that transfer of multiple electrons can be stabilized by screening. 

\textbf{Solutions 13} and \textbf{11} are different again by two charge transfers between Ni hybridized d$_{x^2-y^2}$ and Nd d$_{z^2}$ and $d_{xy}$, but this time with spin flipping. 

\textbf{Solution 14} is unique because of a large change of $\sigma$ along homotopy indicating a continuous change of character. 
Fig.\ref{fig:NDO4} shows the leading NDO between the UHF and scGW limits of \textbf{Solution 14} elucidating the change of character. 
An $\alpha$ electron moves from the hybridized d$_{x^2-y^2}$ orbital on Ni to a hybridized d$_{xy}$ orbital on Nd. 

Fig.~\ref{fig:NDO5} shows NDOs between \textbf{Solutions 15} and \textbf{14}. 
In \textbf{Solution 15}, 1 $\alpha$ electron is moved to d$_{z^2}$ orbital on Nd at the $\Gamma$ point and 1 $\beta$ electron is moved from d$_{z^2}$ on Nd to a hybridized d$_{x^2-y^2}$ on Ni. 
scGW qualitatively changes the character of \textbf{Solution 14}, which is also reflected in the NDO pair 
at the $\Gamma$ point.  
As a result, scGW NDOs consists of the opposite charge movements but at different k-points, which can explain why scGW substantially reduces the energy difference between them in comparison with UHF. 

Finally, almost exactly degenerate \textbf{Solutions 18, 19, } and \textbf{20} are different by many NDOs 
(Fig.~\ref{fig:NDO6}), the structure of which is nearly the same for different k-points. 
The difference between these solutions is cause by a rotation of Ni $d$-orbitals and their hybridization with oxygen's orbitals. 

\section{Conclusions and future work}
We continued our efforts in converging the fully self-consistent Dyson equation for difficult cases. 
We applied the homotopic continuation to the Dyson equation and established the correspondence between 
the UHF and the correlated scGW limits. 
We show that only the real homotopy (or nearly real homotopy with a small $\phi$) makes a physical sense 
since this is the only linear homotopy that can preserve causality. 
We showed usefulness of the developed technique on the example of NdNiO$_2$. 
We observed the following situations:
\begin{itemize}
\item The homotopy connects UHF and scGW limits without a substantial change of character (low $\sigma \approx 0-0.4$).
\item The homotopy connects UHF and scGW limits with a noticeable change of character (intermediate $\sigma \approx 0.4-0.6$).
\item The homotopy connects UHF and scGW limits with a large qualitative change of character ($\sigma \approx 1$).
\item The homotopy cannot be continued toward the scGW limit and the optimization becomes unstable or non-converging. 
\item The homotopy cannot be continued toward the UHF limit and the optimization becomes unstable or non-converging. 
\end{itemize}
Although we did not pursue the goal of converging all solutions, we were able to locate many 
solutions (more than the number of the zero-temperature starting points) 
describing the ground and the excited states.  
We characterized the found solutions by occupation numbers in the k-space as well as through the natural difference orbitals. 
We found that charge transfer is energetically very plausible  and becomes further stabilized for the scGW solutions. 
Our analysis elucidates the important role of d-orbitals on Ni and Nd, between which the charge transfer mainly happens. 
The participation of Nd d-orbitals shows that this compound cannot be illustrated by considering model systems which assume only participation of Ni's d-orbitals and oxygen's orbitals.  
The large participation of Nd also invalidates earlier comparisons\cite{Sawatzky:NdNiO2:model:2020,Alavi:NdNiO2:2022,Sun:NdNiO2:2021} with CaCuO$_2$, where Ca$^{2+}$ is essentially inert. 
The low energy penalty for the charge transfer between different k-points as well as Ni and Nd orbitals qualitatively predicts a substantial conductivity of this compound, which is exactly what has been observed experimentally \cite{Hwang:NdNiO2:conductivity:2020}. 

In the future work, we will explore in detail the influence of the basis set, geometry,  
and the convergence toward the thermodynamic limit since all of these factors 
can influence the energetics of charge transfer.

\section*{Acknowledgments}
This work was supported from the Center for Scalable Predictive Methods for Excitations and Correlated Phenomena (SPEC), which is funded by the U.S. Department of Energy (DOE), Office of Science, Basic Energy Sciences (BES), Chemical Sciences, Geosciences, and Biosciences Division (CSGB), as part of the Computational Chemical Sciences (CCS) program under FWP 70942 at Pacific Northwest National Laboratory (PNNL), a multi-program national laboratory operated for DOE by Battelle. This research also used resources of the National Energy Research Scientific Computing Center, which is supported by the Office of Science of the US Department of Energy under Contract No. DE-AC02-05CH11231 under ERCAP \#0033498: "Benchmarks for NWChem/SPEC libraries"

\section*{Authors Declaration}
The authors have no conflicts to disclose.

\section*{Supplementary Material}
Geometry, k-point occupations for all the solutions, Lowdin occupations of d-orbitals, singular values $\sigma$. 

\section*{Data Availability}
The data supporting the findings of this study are available within the article and its supplementary material.

\renewcommand{\baselinestretch}{1.5}

\clearpage

\end{document}